\documentclass[fleqn,twoside]{article}

\usepackage[headings]{espcrc2}

\usepackage{graphicx}

\usepackage{amsmath}		  
\usepackage{amsfonts}
\usepackage{amssymb}

\newcommand{\msb}{\overline{\mbox{MS}}}

\usepackage[figuresright]{rotating}

\title{Four-Loop Tadpoles: Applications in QCD}

\author{R. Boughezal\address[a]{Institut f\"ur Theoretische Physik und Astrophysik,
    Universit\"at W\"urzburg, \\ Am Hubland, D-97074 W\"urzburg,
    Germany}
  \thanks{This work was supported by the Sofja Kovalevskaja Award of the
    Alexander von Humboldt Foundation sponsored by the German Federal
    Ministry of Education and Research.},
  M. Czakon\addressmark\address{Department of Field Theory and Particle Physics,
    Institute of Physics, \\ University of Silesia, Uniwersytecka 4,
    PL-40007 Katowice, Poland}
  \thanks{Talk given at the 8th DESY Workshop on Elementary Particle Theory, Loops
  and Legs in Quantum Field Theory, April 23-28, 2006, Eisenach, Germany.},
  T. Schutzmeier\addressmark[a]}
       
\begin{document}

\begin{abstract}

  Recent applications of single-scale four-loop tadpoles are briefly
  reviewed. An algorithm for the evaluation of current correlators
  based on differential equations is described and applied to obtain
  high moments of the vacuum polarization function at ${\cal
  O}(\alpha_s^2)$ as a preparation of ${\cal O}(\alpha_s^3)$ calculations.

\end{abstract}

\maketitle

\section{INTRODUCTION}

There are several reasons for the appearance of vacuum integrals in
perturbative calculations. This type of integrals may, in fact, be
introduced for any graph with any number of external legs if large
mass or large momentum expansions are applied. This case is equivalent
to the calculation of Wilson (matching) coefficients as part of the
construction of an Effective Field Theory or to the determination of
vacuum expectation values of composite operators (condensates) in the
Operator Product Expansion. Iterative use of these techniques may
produce integrals where all lines are either massless or carry just
one mass.  Another possibility is the calculation of renormalization
group anomalous dimensions in the $\msb$ scheme. Here, the polynomial
dependence of counterterms on dimensionful parameters is exploited and
the relevant Green functions are first made dimensionless by deriving
with respect to masses and momenta and then evaluated at vanishing
external momenta and with all masses set to unity in order to avoid
spurious infrared divergences. The problem is thus again reduced to
the evaluation of tadpoles, {\it albeit} only to divergent parts.

Recently, problems involving tadpoles at the four-loop level have
become tractable. This impressive progress has been made possible to a
large extent by the Laporta algorithm for the reduction of integrals
to masters described in \cite{Laporta:2001dd}, and by the difference
equation method for the numerical evaluation of the masters proposed
in the same publication. The first sets of integrals have been
evaluated precisely using these principles
\cite{Laporta:2002pg,Schroder:2005va} (see also these proceedings
\cite{Bejdakic:2006vg}) and confirmed by independent
techniques in \cite{Chetyrkin:2006dh}.  At present, other methods are
also available, see \cite{Kniehl:2005yc} and \cite{Chetyrkin:2006bj}.
We should note, that lower precision values, which can be used for
tests, can be obtained by the sector decomposition method
\cite{sectors} and by Mellin-Barnes techniques
\cite{Smirnov:1999gc,Tausk:1999vh} implemented in the {\tt MB} package
\cite{Czakon:2005rk}, see also \cite{Anastasiou:2005cb}.

The first application of four-loop vacuum integrals concerned the
determination of renormalization group parameters, in particular the
four-loop QCD $\beta$-function
\cite{vanRitbergen:1997va,Czakon:2004bu} and the mass anomalous
dimension in the $\msb$ scheme \cite{Vermaseren:1997fq}. As far as the
running of couplings and masses in theories with different mass
scales is concerned, decoupling relations have also been
determined at the same level of precision
\cite{Schroder:2005hy,Chetyrkin:2005ia}.

Another set of applications is related to the vacuum polarization
function, because of its connection to the hadronic production cross
section. In fact, the Taylor expansion coefficients, which are
equivalent through dispersion relations to moments of the cross
section allow one to obtain charm and bottom quark masses with good
precision \cite{Kuhn:2001dm}. Such an analysis has been performed at
four-loops using the first physical moment in
\cite{Boughezal:2006px,Chetyrkin:2006xg} resulting in a large
reduction of the scale dependence.

Four-loop vacuum integrals appear also when QCD corrections to
electroweak observables, in particular the $W$ boson mass
\cite{Awramik:2003rn} and the effective weak mixing angle
\cite{Awramik:2004ge} are computed, through the corrections to the
$\rho$ parameter of Veltman \cite{Veltman:1977kh}. After many
three-loop contributions have been evaluated
\cite{QCD3L,mixed3L,radja}, the pure ${\cal O}(\alpha \alpha_s^3)$
part could finally be done in
\cite{Schroder:2005db,Chetyrkin:2006bj,Boughezal:2006xk}.

Finally, we should mention that the set of solved problems is not
confined to the usual perturbation theory, since observables in high
temperature QCD could also have been evaluated using similar techniques
\cite{Kajantie:2002wa}.

\section{CURRENT CORRELATORS}

One of the computations mentioned in the Introduction concerned the
vacuum polarization function and in particular its Taylor expansion
coefficients. It turned out that the evaluation of the first physical
moment (expansion up to the second order in the external momentum
squared) is already at the edge of the present
techniques. Phenomenology requires, however, further moments to be
computed.

Let us first explain the origin of the difficulties.  Since the
expansion operator is $\partial^2/\partial q_\mu \partial q^\mu$ and
the evaluation point is at the origin, every term of the expansion
introduces two dots (additional powers of the denominators) and one
irreducible numerator. This leads to a steep growth of the number of
integrals to reduce. When using the Laporta algorithm this means
solving systems of equations involving millions of variables already
for the first physical moment, and thus the next moment is in
practice unreachable in any reasonable amount of time. It might seem
that a better algorithm for tadpole reduction could solve this
problem, but the experience made at the three-loop level speaks for
the contrary. In fact, highly optimized software
\cite{Steinhauser:2000ry} used in \cite{Chetyrkin:1996cf} could only
generate seven terms in a calculation that took several months.

Obviously, an algorithm of different complexity is needed
here. Fortunately, the expansion coefficients are not given by
unrelated integrals, but can be generated with linear complexity from
differential equations. In fact, the vacuum polarization function is
expressed through propagator integrals $P_i(z, \epsilon)$, with
$z=q^2/4m^2$ and $d=4-2\epsilon$ the dimension of space-time (a factor
of $m^{-8\epsilon}$ has been taken out). Since a reduction to masters
exists, we have a system of linear equations
\begin{equation}
\frac{d}{d z} P_i(z,\epsilon) = A_{ij}(z,\epsilon) P_j(z,\epsilon),
\end{equation}
where the matrix $A_{ij}(z,\epsilon)$ has a block-triangular form with
elements being rational functions of $z$ and $\epsilon$. The structure
of the matrix allows one to reduce the problem to a solution of
several small systems of differential equations, which are easily
integrated after expansion in both variables. Vacuum integrals are now
only the boundary conditions, and will thus have little dots and/or
irreducible numerators. On the other hand, to a good approximation,
every next term of the Taylor expansion is obtained by repeating the
same procedure within the same amount of time (linear complexity).

\begin{figure*}
\includegraphics[width=15cm]{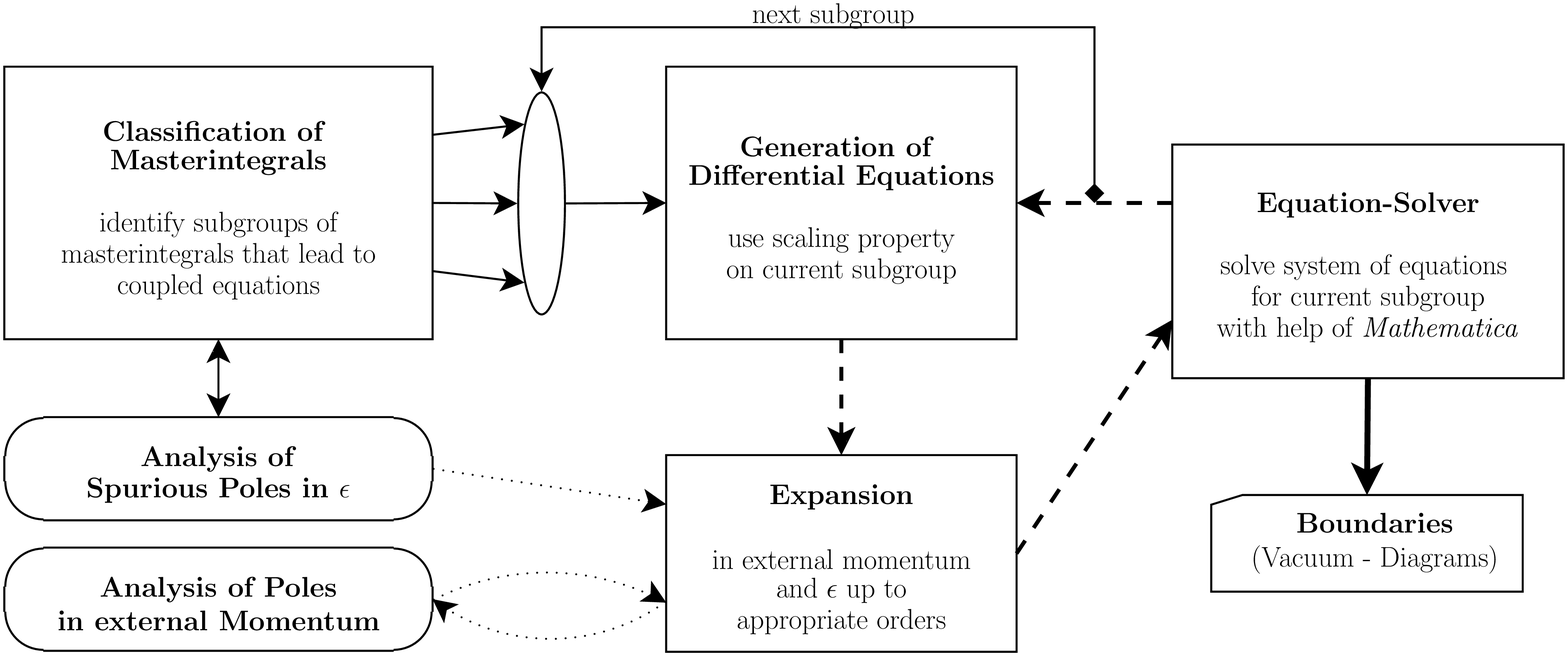}
\caption{\label{schema} Schematic view of an implementation of
  automatic expansions of two-point functions.}
\end{figure*}

There is of course, a price to pay for the improvement of computational
complexity. First of all, what is needed is a reduction of massive
four-loop propagator integrals. Even though, they will be much less
numerous (of the order of $10^4$ for the four-loop vacuum
polarization), they involve two variables. Second, one needs an
automated approach to the treatment of spurious poles in both $z$ and
$\epsilon$. We have constructed such an implementation the sketch of
which can be found in Fig.~\ref{schema}. We should note at this point
that our method is an evolution of the one proposed for the two-loop
sunrise graph in \cite{Caffo:1998du}

As a first application, we have applied our solution to the three-loop
vacuum polarization function. With a prefactor of $(\alpha_s/\pi)^3 \; 3 Q^2/16\pi^2$, we
have obtained the following 30 terms in a matter of hours (zeroth
order term is zero)

\begin{align*}
\hspace*{-2.4em}+\, C_F^2& \; (0.15413 \,z + 0.30248 \,z^{2} \\
\hspace*{-2.4em}+ \,& 0.27107 \,z^{3}  + 0.27851 \,z^{4} + 0.34237 \,z^{5} \\
\hspace*{-2.4em}+ \,& 0.45346 \,z^{6}  + 0.60069 \,z^{7} + 0.77496 \,z^{8} \\
\hspace*{-2.4em}+ \,& 0.96941 \,z^{9}   + 1.1789 \,z^{10} + 1.3996 \,z^{11}\\
\hspace*{-2.4em}+ \,& 1.6287 \,z^{12} + 1.8639 \,z^{13} + 2.1036 \,z^{14}\\
\hspace*{-2.4em}+ \,& 2.3465 \,z^{15} + 2.5917 \,z^{16} + 2.8382 \,z^{17}\\
\hspace*{-2.4em}+ \,& 3.0855 \,z^{18}  + 3.3331 \,z^{19} + 3.5805 \,z^{20}\\
\hspace*{-2.4em}+ \,& 3.8276 \,z^{21} + 4.074 \,z^{22} + 4.3195 \,z^{23}\\
\hspace*{-2.4em}+ \,& 4.5639 \,z^{24} + 4.8072 \,z^{25} + 5.0492 \,z^{26}\\
\hspace*{-2.4em}+ \,& 5.2899 \,z^{27} + 4.5292 \,z^{28} + 5.767 \,z^{29}\\
\hspace*{-2.4em}+ \,& 6.0034 \,z^{30})
\end{align*}

\begin{align*}
\hspace*{-2.4em}-\,C_A&\,C_F\;(0.0069711 \,z - 0.22146 \,z^2 \\
\hspace*{-2.4em}+ \,& 0.015317 \,z^3  + 1.22184\,z^4 + 0.37748 \,z^5 \\
\hspace*{-2.4em}+ \,& 0.49364 \,z^6 + 0.58131 \,z^7 +0.64841 \,z^8 \\
\hspace*{-2.4em}+ \,& 0.70043 \,z^9 + 0.74121 \,z^{10} + 0.77344 \,z^{11}\\
\hspace*{-2.4em}+ \,& 0.79907 \,z^{12} + 0.81954 \,z^{13} + 0.83592 \,z^{14}\\
\end{align*}

\begin{align*}
\hspace*{-2.4em}+ \,& 0.84902 \,z^{15} + 0.85946 \,z^{16} + 0.86774 \,z^{17}\\
\hspace*{-2.4em}+ \,& 0.87423 \,z^{18} + 0.87925 \,z^{19} + 0.88304 \,z^{20}\\
\hspace*{-2.4em}+ \,& 0.88579 \,z^{21} + 0.88768 \,z^{22} + 0.88884 \,z^{23}\\
\hspace*{-2.4em}+ \,& 0.88938 \,z^{24} + 0.88939 \,z^{25} + 0.88894 \,z^{26}\\
\hspace*{-2.4em}+ \,& 0.88812 \,z^{27} + 0.88696 \,z^{28} + 0.88553 \,z^{29}\\
\hspace*{-2.4em}+ \,& 0.88385 \,z^{30})
\end{align*}

\begin{align*}
\hspace*{-2.4em}+\, C_F&\,T_F\;(-2.39562 \,z - 0.016957 \,z^{2}\\ 
\hspace*{-2.4em}- \,& 0.10262 \,z^{3}  -  0.13008 \,z^{4} - 0.13959 \,z^{5} \\
\hspace*{-2.4em}- \,& 0.14211 \,z^{6}  -  0.14155 \,z^{7} - 0.13957 \,z^{8} \\
\hspace*{-2.4em}- \,& 0.13694 \,z^{9}  -  0.13403 \,z^{10} - 0.13105 \,z^{11}\\
\hspace*{-2.4em}- \,& 0.12809 \,z^{12} -  0.12522 \,z^{13} - 0.12246 \,z^{14}\\
\hspace*{-2.4em}- \,& 0.11982 \,z^{15} -  0.11731 \,z^{16} - 0.11492 \,z^{17}\\
\hspace*{-2.4em}- \,& 0.11264 \,z^{18} -  0.11048 \,z^{19} - 0.10843 \,z^{20}\\
\hspace*{-2.4em}- \,& 0.10647 \,z^{21} -  0.10461 \,z^{22} - 0.10284 \,z^{23}\\
\hspace*{-2.4em}- \,& 0.10114 \,z^{24} -  0.099528 \,z^{25} - 0.097983 \,z^{26}\\
\hspace*{-2.4em}- \,& 0.096504 \,z^{27} -  0.095088 \,z^{28} - 0.09373 \,z^{29}\\
\hspace*{-2.4em}- \,& 0.092427 \,z^{30})
\end{align*}

\begin{align*}
\hspace*{-2.4em}+ \,C_F&\, T_F\, n_l\,(-1.99342 \,z + 0.68237 \,z^{2} \\
\hspace*{-2.4em}+ \,& 0.64329 \,z^{3}  + 0.63725 \,z^{4} + 0.63623 \,z^{5} \\
\hspace*{-2.4em}+ \,& 0.63507 \,z^{6}  + -1.63287 \,z^{7} + 0.62966 \,z^{8} \\
\hspace*{-2.4em}+ \,& 0.62566 \,z^{9}  + 0.6211 \,z^{10} + 0.61616 \,z^{11}\\
\hspace*{-2.4em}+ \,& 0.61097 \,z^{12} + 0.60564 \,z^{13} + 0.60024 \,z^{14}\\
\hspace*{-2.4em}+ \,& 0.59482 \,z^{15} + 0.58943 \,z^{16} + 0.58409 \,z^{17}\\
\hspace*{-2.4em}+ \,& 0.57883 \,z^{18} + 0.57365 \,z^{19} + 0.56858 \,z^{20}\\
\hspace*{-2.4em}+ \,& 0.5636 \,z^{21}  + 0.55874 \,z^{22} + 0.55398 \,z^{23}\\
\hspace*{-2.4em}+ \,& 0.54933 \,z^{24} + 0.54479 \,z^{25} + 0.54035 \,z^{26}\\
\hspace*{-2.4em}+ \,& 0.53603 \,z^{27} + 0.5318 \,z^{28} + 0.52768 \,z^{29}\\
\hspace*{-2.4em}+ \,& 0.52365 \,z^{30}) 
\end{align*}

Let us note finally that the most costly part of the calculation is
not the preparation of differential equation or their solution, but
the reduction of the integrals which occur in the actual diagrams. The
reduction at the four-loop level is under way.

\section{CONCLUSIONS}

We have reviewed the current status of calculations involving
four-loop tadpoles. Further progress will depend crucially on the development
of more efficient algorithms of reduction of integrals to masters. We
have made a first step in this direction and presented a method to
evaluate high order expansions of current-current correlators. As an
illustration, we have applied this approach to the vacuum polarization
function at the three-loop level.

\end{document}